\documentclass[reqno]{amsart}
\usepackage[english]{babel}            \usepackage[latin1]{inputenc}
\usepackage{latexsym,amsfonts}         \usepackage{graphics}
\usepackage{ulem}                      \usepackage{hhline}
\usepackage{dsfont}                    \usepackage{mathrsfs}
\usepackage{mathpazo}
        \usepackage{euscript}
              \usepackage[pdftex]{hyperref}
\makeatletter\@addtoreset{equation}{section}

\newtheorem {theorem}{Theorem}[section]
\newtheorem {definition}[theorem]{Definition}
\newtheorem {remark}[theorem]{Remark}
\newtheorem {proposition}[theorem]{Proposition}

\newcommand{\C}{\mathbb C} \newcommand{\R}{\mathbb R}

\newcommand{\fin}{\hfill $\square$}

\begin{document}

\title{A family of circular Bargmann transforms}
\author{Zouha\"{i}r MOUAYN}
\address{Department of Mathematics, Faculty of Sciences and Technics (M'Ghila)
         \newline
         Sultan Moulay Slimane University, BP 523, B\'{e}ni Mellal, Morocco}
\email{mouayn@gmail.com}

 \maketitle

\begin{abstract}
When considering a charged particle evolving in the Poincar\'{e} disk under
influence of a uniform magnetic field with a strength proportional to\ $%
\gamma +1,$ we construct for all hyperbolic Landau level $\epsilon
_{m}^{\gamma }=4m\left( \gamma -m\right) ,m\in \mathbb{Z}_{+}\cap \left[
0,\gamma /2\right] $ a \ family of coherent states transforms labeled by ($%
\gamma ,m)$ and mapping isometrically square integrable functions on the
unit circle with respect to the measure $\sin ^{\gamma -2m}\left( \theta
/2\right) d\theta $ onto spaces of bound states of the particle. These
transforms are called circular Bargmann transforms.
\end{abstract}

\section{Introduction}

\noindent\ The classical Bargmann transform \cite{1} can be defined
as
\begin{equation}
\mathcal{B}^{H}\left[ \varphi \right] (z) := \pi ^{-\frac{1}{4}%
}e^{-\frac{1}{2}z^{2}}\int_{\R}e^{-\xi ^{2}+2\xi z}\varphi \left(
\xi \right) d\xi , \quad  z\in \C.  \label{1.1}
\end{equation}
It maps isometrically the space $L^{2}\left( \R,d\xi \right) $ of
square integrable functions on the real line onto the Fock space
$\frak{F} \left( \C\right) $ of entire Gaussian square-integrable
functions. Since the involved kernel in \eqref{1.1} is related to
the generating function of Hermite polynomials \cite{2}, we have
used the notation $\mathcal{B}^{H}$.  In the same paper
(\cite[p.203]{1}) V. Bargmann has also introduced a second transform
labeled by a parameter $\gamma >0$ as
\begin{equation}
\mathcal{B}_{\gamma }^{L}:L^{2}\left(\R_{+},\frac{x^{\gamma }}{\Gamma
\left( \gamma +1\right) }dx\right)\rightarrow \mathcal{A}^{\gamma }\left( \mathbb{D%
}\right)   \label{1.3}
\end{equation}
defined by
\begin{equation}
\mathcal{B}_{\gamma }^{L}\left[ \psi \right] (z) :=\frac{\left(
\frac{\gamma }{\pi }\right) ^{\frac{1}{2}}}{\left( 1-z\right) ^{\gamma +1}}%
\int_{0}^{+\infty }\exp \left( -\frac{x}{2}\left( \frac{1+z}{1-z}%
\right) \right) \psi \left( x\right) \frac{x^{\gamma }}{\Gamma \left( \gamma
+1\right) }dx,  \label{1.4}
\end{equation}
where
\begin{equation}
\mathcal{A}^{\gamma }\left( \mathbb{D}\right) :=\left\{ \psi \text{
analytic on }\mathbb{D},\int_{\mathbb{D}}\left| \psi (z) \right|
^{2}\left( 1-\left| z\right| ^{2}\right) ^{\gamma -1}d\mu (z)
<+\infty \right\}   \label{1.5}
\end{equation}
denotes the weighted Bergman space on the unit disk
$\mathbb{D}=\left\{ z\in \C;\left| z\right| <1\right\} $ and
$d\mu(z) $ being the Lebesgue measure on it. The involved kernel
function in \eqref{1.3}  corresponds to the generating function of
Laguerre polynomials \cite{2}. This explains the notation
$\mathcal{B}_{\gamma }^{L}$.

In the present work, we first propose, if it does not exist in the
literature, the following integral transform
\begin{equation}
\mathcal{B}_{\gamma }^{J}:L^{2}(\mathbf{S}^{1},d\sigma _{\gamma
})\rightarrow \mathcal{A}^{\gamma }\left( \mathbb{D}\right)   \label{1.6}
\end{equation}
defined by
\begin{equation}
\mathcal{B}_{\gamma }^{J}\left[ \phi \right] (z) :=\frac{\left(
\frac{\gamma }{\pi }\right) ^{\frac{1}{2}}}{\left( 1-z\right) ^{\frac{1}{2}%
\gamma }}\int_{\mathbf{0}}^{2\pi }\frac{1}{\left( 1-ze^{i\theta
}\right) ^{1+\frac{1}{2}\gamma }}\phi \left( e^{i\theta }\right)
d\sigma _{\gamma }\left( \theta \right) ,  \label{1.7}
\end{equation}
and mapping isometrically the square integrable functions on the
unit circle $\mathbf{S}^{1}\mathbb{=}\left\{ \omega \in
\C\text{,}\left| \omega
\right| =1\right\} $ endowed with $d\sigma _{\gamma }\left( \theta \right) :=%
\frac{2^{\gamma }\Gamma ^{2}\left( \gamma /2+1\right) }{\Gamma
\left( \gamma +1\right) }\sin ^{\gamma }\left( \theta /2\right)
d\theta $ as measure, onto the Bergman space in \eqref{1.5}. We
obtain the kernel function in \eqref{1.7}  by using a generating
function for Jacobi polynomials due to H.M. Srivastava \cite{3}.  Here, we have used the notation $%
\mathcal{B}_{\gamma }^{J}$ ($J$: Jacobi). We also propose a
generalization of the transform in \eqref{1.7}  by replacing the
arrival space in \eqref{1.6}  by the eigenspace (\cite{4}):
\begin{equation}
\mathcal{A}_{m}^{\gamma }\left( \mathbb{D}\right) :=\left\{ \psi :\mathbb{D}%
\rightarrow \C,\Delta _{\gamma }\psi =\epsilon _{m}^{\gamma }\psi
,\int_{\mathbb{D}}\left| \psi (z) \right| ^{2}\left( 1-\left|
z\right| ^{2}\right) ^{\gamma -1}d\mu (z) <+\infty \right\}
\label{1.8}
\end{equation}
of the second order differential operator
\begin{equation}
\Delta _{\gamma }:=-4\left( 1-\left| z\right| ^{2}\right) \left( \left(
1-\left| z\right| ^{2}\right) \frac{\partial ^{2}}{\partial z\partial
\overline{z}}-\left( \gamma +1\right) \overline{z}\frac{\partial }{\partial
\overline{z}}\right)   \label{1.9}
\end{equation}
with the eigenvalue (\textit{hyperbolic Landau level}):
\begin{equation}
\epsilon _{m}^{\gamma }:=4m\left( \gamma -m\right) , \quad m=0,1,2,\cdots ,\left[ \frac{%
\gamma }{2}\right] ,  \label{1.10}
\end{equation}
where $\left[ x\right] $ denotes the greatest integer less than $x$.
The operator in \eqref{1.9}  can be unitarly intertwined to
represent the Schr\"{o}dinger operator of a charged particle evolving in the Poincar%
\'{e} disk under influence of a uniform magnetic field with a strength
proportional to\ ($\gamma +1$)$.$ For $m=0,$ the space $\mathcal{A}%
_{0}^{\gamma }\left( \mathbb{D}\right) $ in (1.8) coincides with the
Bergman space $\mathcal{A}^{\gamma }\left( \mathbb{D}\right) $ in
\eqref{1.5}. For $m\neq 0,$ we precisely construct the integral
transform
\begin{equation}
\mathcal{B}_{\gamma ,m}^{J}:L^{2}(\mathbf{S}^{1},d\sigma _{\left( \gamma
-2m\right) })\rightarrow \mathcal{A}_{m}^{\gamma }\left( \mathbb{D}\right)
\label{1.11}
\end{equation}
defined by
\begin{eqnarray}
\mathcal{B}_{\gamma ,m}^{J}\left[ \phi \right] (z)  &=&\frac{%
\left( \frac{\Gamma \left( \gamma +1-m\right) }{\pi m!\Gamma \left( \gamma
-2m\right) }\right) ^{\frac{1}{2}}}{\left( 1-z\right) ^{\frac{\gamma }{2}}}%
\int_{0}^{2\pi }\frac{1}{\left( 1-e^{i\theta }z\right)
^{\frac{\gamma }{2}+1}}\left( \frac{\left( \overline{z}-1\right)
\left( 1-e^{i\theta }z\right) }{\left( 1-\left| z\right| ^{2}\right)
}\right) ^{m}  \label{1.12}
\\
&&\times {_2F_1}\left(
\begin{array}{c}
-m,\frac{\gamma }{2}-m+1 \\
1+\gamma -2m
\end{array}
\bigg | \frac{\left( 1-e^{i\theta }\right) \left( 1-\left| z\right| ^{2}\right)
}{\left( 1-\overline{z}\right) \left( 1-e^{i\theta }z\right) }\right) \phi
\left( e^{i\theta }\right) d\sigma _{\left( \gamma -2m\right) }\left( \theta
\right), \nonumber
\end{eqnarray}
where ${_2F_1}\left( -m,.,.\mid \cdot \right) $ is a
hypergeometric function which can be written in terms of the Jacobi
polynomial \cite{2}.  We obtain the kernel function in \eqref{1.12}
by using a bilateral generating function for Gauss hypergeometric
sums due to L.Weisner \cite{5}.  Our method in constructing the
transforms \eqref{1.7}  and \eqref{1.12}  is based on coherent
states analysis.

The paper is organized as follows. In Section 2, we recall briefly the
coherent states formalism we will be using. This formalism is applied in
Sections 3 and 4 \ so as to establish the announced transforms.

\section{Coherent states}

Following (\cite[pp.72-76]{5}), we give a summary of the coherent
states formalism we will be using. Let $(X,\nu )$\ be a measure
space and let $\mathcal{A}\subset L^{2}(X,\nu )$\ be a closed
subspace of infinite dimension. Let $\left\{ \Phi _{n}\right\}
_{n=0}^{\infty }$ be an orthogonal basis of $\mathcal{A}$
satisfying, for arbitrary $\xi \in X,$
\begin{equation}
\kappa \left( \xi \right) :=\sum_{n=0}^{\infty }\rho _{n}^{-1}\left| \Phi
_{n}\left( \xi \right) \right| ^{2}<+\infty ,  \label{2.1}
\end{equation}
where $\rho _{n}:=\left\| \Phi _{n}\right\| _{L^{2}(X)}^{2}$\ .
Define
\begin{equation}
\mathcal{K}(\xi ,\zeta ):=\sum_{n=0}^{\infty }\rho _{n}^{-1}\Phi _{n}\left(
\xi \right) \overline{\Phi _{n}(\zeta )},\quad \xi ,\zeta \in X.  \label{2.2}
\end{equation}
Then, $\mathcal{K}(\xi ,\zeta )$\ is a reproducing kernel, $\mathcal{A}$ is
the corresponding reproducing kernel Hilbert space and $\kappa \left( \xi
\right) =$ $\mathcal{K}(\xi ,\xi )$. Let $\mathcal{H}$ be another Hilbert
space with $\dim \mathcal{H}=\infty $ and $\left\{ \phi _{n}\right\}
_{n=0}^{\infty }$\ be an orthonormal basis of $\mathcal{H}.$ Therefore$,$
define a coherent state as a ket vector $\mid \xi >\in \mathcal{H}$ labelled
by a point $\xi \in X$ as
\begin{equation}
\mid \xi >:=\left( \mathcal{K}(\xi ,\xi )\right) ^{-\frac{1}{2}%
}\sum_{n=0}^{\infty }\frac{\Phi _{n}\left( \xi \right) }{\sqrt{\rho _{n}}}%
\mid \phi _{n}>.  \label{2.3}
\end{equation}
We rewrite \eqref{2.3}  using Dirac's bra-ket notation as
\begin{equation}
<x\mid \xi >=\left( \mathcal{K}(\xi ,\xi )\right) ^{-\frac{1}{2}%
}\sum_{n=0}^{\infty }\frac{\Phi _{n}\left( \xi \right) }{\sqrt{\rho _{n}}}%
<x\mid \phi _{n}>.  \label{2.4}
\end{equation}
By definition, it is straightforward to show that $<\xi \mid \xi >=1$\ and
the coherent state transform $W:\mathcal{H}\rightarrow \mathcal{A}\subset
L^{2}(X,\nu )$ defined by
\begin{equation}
W\left[ \phi \right] \left( \xi \right) :=\left( \mathcal{K}(\xi ,\xi
)\right) ^{\frac{1}{2}}<\xi \mid \phi >  \label{2.5}
\end{equation}
is an isometry. Thus, for $\phi ,\psi \in \mathcal{H}$, we have
\begin{equation}
<\phi \mid \psi >_{\mathcal{H}}=<W\left[ \phi \right] \mid W\left[ \psi %
\right] >_{L^{2}\left( X\right) }=\int_{X}d\nu \left( \xi \right)
\mathcal{K}(\xi ,\xi )<\phi \mid \xi ><\xi \mid \psi >  \label{2.6}
\end{equation}
and thereby we have a resolution of the identity
\begin{equation}
\mathbf{1}_{\mathcal{H}}=\int_{X}d\nu \left( \xi \right) \mathcal{K}%
(\xi ,\xi )\mid \xi ><\xi \mid ,  \label{2.7}
\end{equation}
where $\mathcal{K}(\xi ,\xi )$ appears as a weight function.

\section{The transform $\mathcal{B}_{\protect\gamma }^{J}$}

According to the above formalism, we define coherent states with the
following elements:

\begin{itemize}
\item[$\bullet$] $(X,\nu ):=(\mathbb{D},\left( 1-\left| z\right|
^{2}\right) ^{\gamma -1}d\mu (z) )$ with $\gamma >0$

\item[$\bullet$] $\mathcal{A}:=\mathcal{A}^{\gamma }\left( \mathbb{D}%
\right) $ \ is the weighted Bergman space in \eqref{1.5}

\item[$\bullet$] A well known orthonormal basis of $\mathcal{A}$ has the form
\begin{equation}
\Phi _{n}^{\gamma }(z) :=\left( \frac{\gamma \Gamma \left(
\gamma +1+n\right) }{\pi \Gamma \left( \gamma +1\right) n!}\right) ^{\frac{1%
}{2}}z^{n}, \quad n=0,1,2,\cdots .  \label{3.1}
\end{equation}

\item[$\bullet$] The diagonal function of the reproducing kernel of $\mathcal{A}$ is
given by
\begin{equation}
K_{\gamma }\left( z,z\right) \mathbf{\ }=\frac{\gamma }{\pi }\left( 1-\left|
z\right| ^{2}\right) ^{-\gamma -1}, \quad z\in \mathbb{D.}  \label{3.2}
\end{equation}

\item[$\bullet$] $\mathcal{H}$:$=L^{2}(\mathbf{S}^{1},d\sigma
_{\gamma })$ is the space carrying the coherent states, which is
endowed with the measure (\cite{7}):
\begin{equation}
d\sigma _{\gamma }\left( \theta \right) :=\frac{2^{\gamma }\Gamma ^{2}\left(
\frac{\gamma }{2}+1\right) }{\Gamma \left( \gamma +1\right) }\left( \sin
\frac{\theta }{2}\right) ^{\gamma }\frac{d\theta }{2\pi }.  \label{3.4}
\end{equation}

 \item[$\bullet$] A basis of $\mathcal{H}$ is consisting of
circular Jacobi polynomials (\cite{7,8,9}) given by
\begin{equation}
g_{n}^{\gamma }\left( e^{i\theta }\right) :=\frac{\left( \gamma +1\right)
_{n}}{n!}\; {_2F_1}\left( -n,\frac{\gamma }{2}+1,\gamma +1  | 1-e^{i\theta
}\right) ,  \label{3.5}
\end{equation}
where the notation in \cite{8} is adapted. The ket vectors we shall
take are the orthonormalized functions:
\begin{equation}
\mid n;\gamma >:=\frac{\sqrt{n!}}{\sqrt{\left( \gamma +1\right) _{n}}}%
g_{n}^{\gamma }  \label{3.6}
\end{equation}
\end{itemize}

\begin{definition} \label{Def3.1}
For $\gamma >0$, coherent states belonging to
$L^{2}(\mathbf{S}^{1},d\sigma _{\gamma })$ are defined according to
\eqref{2.3} by the series expansion
\begin{equation}
\mid z;\gamma >:=\left( K_{\gamma }\left( z,z\right) \right) ^{-\frac{1}{2}%
}\sum_{n=0}^{+\infty }\Phi _{n}^{\gamma }(z) \mid n;\gamma > ;
\label{3.7}
\end{equation}
where $z\in \mathbb{D}$ are labeling points.
\end{definition}

We now give a closed form for these coherent states.

\begin{proposition} \label{Prop3.1} For  $\gamma >0,$  the
wave functions of the states in \eqref{3.7} are of the form
\begin{equation}
<e^{i\theta }\mid z;\gamma >=\left( 1-\left| z\right| ^{2}\right) ^{\frac{%
1+\gamma }{2}}\left( 1-z\right) ^{-\frac{1}{2}\gamma }\left(
1-ze^{i\theta }\right) ^{-1-\frac{1}{2}\gamma } , \label{3.8}
\end{equation}
 where  $z\in \mathbb{D}$  is a fixed labeling point and  $%
e^{i\theta } \in \mathbf{S}^{1}.$
\end{proposition}

\noindent{\it  Proof.}  We start by replacing the three pieces in
\eqref{3.7} by their expressions respectively in \eqref{3.1},
\eqref{3.2} and \eqref{3.6}. We get successively
\begin{align}
<e^{i\theta }\mid z;\gamma >&=\sqrt{\pi }\left( 1-\left| z\right|
^{2}\right) ^{\frac{1}{2}\gamma +\frac{1}{2}}\sum_{n=0}^{+\infty
}\Phi _{n}^{\gamma }(z) \mid n;\gamma >  \label{3.9}
\\
&=\sqrt{\pi }\left( 1-\left| z\right| ^{2}\right) ^{\frac{1}{2}\gamma +\frac{1%
}{2}}\sum_{n=0}^{+\infty }\left( \frac{\Gamma \left( \gamma +1+n\right) }{%
\pi \Gamma \left( \gamma +1\right) n!}\right) ^{\frac{1}{2}}z^{n}\mid
n;\gamma >  \label{3.10}
\\
&=\left( 1-\left| z\right| ^{2}\right) ^{\frac{\gamma +1}{2}%
}\sum_{n=0}^{+\infty }\frac{\sqrt{\left( \gamma +1\right) _{n}}}{\sqrt{n!}}%
z^{n}\mid n;\gamma >  \label{3.11}
\\
&=\left( 1-\left| z\right| ^{2}\right) ^{\frac{\gamma +1}{2}%
}\sum_{n=0}^{+\infty }\frac{\left( \gamma +1\right) _{n}}{n!}%
\; {_2F_1}\left( -n,\frac{\gamma }{2}+1,\gamma +1 \big | 1-e^{i\theta
}\right) z^{n}. \label{3.12}
\end{align}
Now, we write the hypergeometric function in \eqref{3.12} in terms
of the Jacobi polynomial via the relation (\cite[p.999]{2}):
\begin{equation}
P_{n}^{\left( \alpha ,\beta -n\right) }\left( u\right) =\left(
\begin{array}{c}
n+\alpha  \\
n
\end{array}
\right) \left( \frac{1+u}{2}\right) ^{n}\; {_2F_1}\left( -n,-\beta ,\alpha
+1\bigg | \frac{u-1}{u+1}\right)   \label{3.13}
\end{equation}
for the parameters $\alpha =\gamma ,$ $1+\gamma /2=-\beta $ and $%
u=2e^{-i\theta }-1.$ We obtain that
\begin{equation}
{_2F_1}\left( -n,\frac{\gamma }{2}+1,\gamma +1\big | 1-e^{i\theta }\right) =%
\frac{n!}{\left( \gamma +1\right) _{n}}e^{in\theta }P_{n}^{\left( \gamma
,-\left( 1+\frac{\gamma }{2}\right) -n\right) }\left( 2e^{-i\theta
}-1\right)   \label{3.14}
\end{equation}
Next, inserting \eqref{3.14} into \eqref{3.12}, gives that
\begin{equation}
<e^{i\theta }\mid z;\gamma >=\left( 1-\left| z\right| ^{2}\right) ^{\frac{%
\gamma +1}{2}}\sum_{n=0}^{+\infty }\left( ze^{i\theta }\right)
^{n}P_{n}^{\left( \gamma ,-\left( 1+\frac{\gamma }{2}\right) -n\right)
}\left( 2e^{-i\theta }-1\right)   \label{3.15}
\end{equation}
Now, we make use of the generating formula (\cite[p.154]{3}):
\begin{eqnarray}
\sum_{n=0}^{+\infty }\left(
\begin{array}{c}
n+\nu  \\
n
\end{array}
\right) P_{n+\nu }^{\left( \alpha ,\beta -n\right) }\left( u\right) t^{n}
&=&\left( 1-t\right) ^{\beta }\left( 1-\frac{1}{2}\left( 1+u\right) t\right)
^{-\alpha -\beta -\nu -1}  \label{3.17} \\
&&\times P_{\nu }^{\left( \alpha ,\beta \right) }\left( \frac{\left( u-\frac{%
1}{2}\left( 1+u\right) t\right) }{\left( 1-\frac{1}{2}\left(
1+u\right) t\right) }\right)   \nonumber
\end{eqnarray}
for $\nu =0,$ $\alpha =\gamma ,$ $\beta =-\left( 1+\frac{\gamma
}{2}\right) , $ $u=2e^{-i\theta }-1$ and $t=ze^{i\theta }.$ We
arrive at the expression  \eqref{3.8}. This ends the proof.  \fin \\

Now, let $\phi \in L^{2}(\mathbf{S}^{1},d\sigma _{\gamma })$. Using
the formalism in Section 2, we define
\begin{align}
\mathcal{B}_{\gamma }^{J}\left[ \phi \right] (z) &=\left( K_{\gamma
}\left( z,z\right) \right) ^{\frac{1}{2}}\left\langle \phi ,\mid
z;\gamma >\right\rangle _{\mathcal{H}}  \label{3.18} \\ &=\left(
\frac{\gamma }{\pi }\left( 1-\left| z\right| ^{2}\right) ^{-\gamma
-1}\right) ^{\frac{1}{2}}\int_{0}^{2\pi }<e^{i\theta }\mid z;\gamma
>\phi \left( e^{i\theta }\right) d\sigma _{\gamma }\left( \theta
\right).   \label{3.19}
\end{align}
Next, using Proposition \eqref{3.1}  we can state the following
 result.

\begin{theorem} \label{Thm3.1}  Let  $\gamma >0.$  Then, the
coherent states transform associated with the coherent states in
\eqref{3.7} is\ the\ isometry  $\mathcal{B}_{\gamma
}^{J}:L^{2}(\mathbf{S}^{1},d\sigma _{\gamma })\rightarrow A^{\gamma
} (\mathbb{D})$   given by
\begin{equation}
\mathcal{B}_{\gamma }^{J}\left[ \phi \right] (z) =\frac{\left(
\frac{\gamma }{\pi }\right) ^{\frac{1}{2}}}{\left( 1-z\right) ^{\frac{1}{2}%
\gamma }}\int_{0}^{2\pi }\frac{1}{\left( 1-ze^{i\theta }\right) ^{1+%
\frac{1}{2}\gamma }}\overline{\phi \left( e^{i\theta }\right)
}d\sigma _{\gamma }\left( \theta \right) .  \label{3.20}
\end{equation}
\end{theorem}

\begin{definition} \label{Def3.11}
  The coherent state transform in \eqref{3.20} will be called the circular Bargmann transform of
attached to the lowest hyperbolic Landau level.
\end{definition}

\section{The transform $\mathcal{B}_{\protect\gamma ,m}^{J}$}

As in Section 3, the elements we will be using to construct coherent
states are as follows:

\begin{itemize}

\item[$\bullet$] $(X,\nu )=(\mathbb{D},\left( 1-\left| z\right|
^{2}\right) ^{\gamma -1}d\mu (z) ),\gamma >0$.

\item[$\bullet$] $\mathcal{A}:=\mathcal{A}_{m}^{\gamma }\left( \mathbb{D}%
\right) $ \ is the eigenspace in \eqref{1.8}.

\item[$\bullet$] An orthonormal basis of $\mathcal{A}$ is given by
(\cite[p.3, Eq.(2.9)]{4}, with $\gamma =2\nu -1$):
\begin{eqnarray}
\Phi _{n}^{\gamma ,m}(z)  &=&\left( -1\right) ^{n}\left( \frac{%
\gamma -2m}{\pi }\right) ^{\frac{1}{2}}\left( \frac{n!\Gamma \left( \gamma
-m+1\right) }{m!\Gamma \left( \gamma -2m+n+1\right) }\right) ^{\frac{1}{2}}
\label{4.1} \\
&&\times \left( 1-\left| z\right| ^{2}\right) ^{-m}\overline{z}%
^{m-n}P_{n}^{\left( m-n,\gamma -2m\right) }\left( 1-2\left| z\right|
^{2}\right)  .\nonumber
\end{eqnarray}

\item[$\bullet$] The diagonal function of the reproducing kernel of
$\mathcal{A}$ \ is given by (\cite[p.3]{4}):
\begin{equation}
K_{\gamma ,m}\left( z,z\right) \mathbf{\ }=\frac{(\gamma -2m)}{\pi }\left(
1-\left| z\right| ^{2}\right) ^{-1-\gamma }.  \label{4.2}
\end{equation}

\item[$\bullet$] $\mathcal{H}$:=$L^{2}(\mathbf{S}^{1},d\sigma _{\gamma
^{\prime }})$ is the Hilbert space carrying the coherent states with the
parameter $\gamma ^{\prime }:=\gamma -2m>0,$ where $\gamma $ is the fixed
parameter in the definition of the space $\mathcal{A}_{m}^{\gamma }\left(
\mathbb{D}\right) .$

\item[$\bullet$]  The ket vectors we take are the same
orthonormalized functions in \eqref{3.4}  but now depending on the
parameter $\gamma ^{\prime }$ as
\begin{equation}
\mid n;\gamma ^{\prime }>:=\frac{\sqrt{n!}}{\sqrt{\left( \gamma ^{\prime
}+1\right) _{n}}}g_{n}^{\gamma ^{\prime }}.  \label{4.3}
\end{equation}
\end{itemize}

\begin{definition} \label{Def4.1}   For  $\gamma >0$ and $m=0,1,\cdots\left[
\frac{\gamma }{2}\right] ,$  the coherent states\ belonging to  $%
L^{2}(\mathbf{S}^{1},d\sigma _{\left( \gamma -2m\right) })\mathcal{\ }$%
  are defined according to \eqref{2.3} by the series
expansion
\begin{equation}
\mid z;\gamma ,m>:=\left( K_{\gamma ,m}\left( z,z\right) \right) ^{-\frac{1}{%
2}}\sum_{n=0}^{+\infty }\Phi _{n}^{\gamma ,m}(z) \mid n;(\gamma
-2m)> , \label{4.4}
\end{equation}
 where  $z\in \mathbb{D}$ are labeling points.
 \end{definition}
We should note that coherent states attached to hyperbolic Landau
levels with similar form \eqref{4.4} have been performed in
\cite{10} and \cite{4} but with different choices for the Hilbert spaces $%
\mathcal{H}$ carrying them. Here the space $\mathcal{H}$ is $L^{2}(\mathbf{S}%
^{1}, d\sigma _{\left( \gamma -2m\right) })$ spanned by the ket
vectors \eqref{4.3} . We now give a closed form for these coherent
states in \eqref{4.4}.

\begin{proposition} \label{Prop4.1}   The wave functions of the states
in \eqref{4.4} can be expressed in a closed form as
\begin{align}
<e^{i\theta }\mid z,\gamma ,m>&=\left( \frac{\Gamma \left( \gamma
-m+1\right) }{m!\Gamma \left( \gamma -2m+1\right) }\right) ^{\frac{1}{2}}%
\frac{\left( 1-\left| z\right| ^{2}\right) ^{\frac{\gamma +1}{2}}}{\left(
1-z\right) ^{\frac{\gamma }{2}}\left( 1-ze^{i\theta }\right) ^{\frac{\gamma
}{2}+1}}  \label{4.5} \\
&\times \left( \frac{\left( \overline{z}-1\right) \left(
1-ze^{i\theta }\right) }{\left( 1-\left| z\right| ^{2}\right)
}\right) ^{m} {_2F_1}\left(
\begin{array}{c}
-m,\frac{\gamma }{2}-m+1 \\
1+\gamma -2m
\end{array}
\bigg |  \frac{\left( 1-\left| z\right| ^{2}\right) \left( 1-e^{i\theta
}\right) }{\left( 1-\overline{z}\right) \left( 1-ze^{i\theta
}\right) }\right), \nonumber
\end{align}
 where  $z\in \mathbb{D}$  is a fixed labeling point and  $%
e^{i\theta }\in \mathbf{S}^{1}.$
\end{proposition}

\noindent{\it  Proof.}  We start from \eqref{4.4} by replacing the
three pieces by their expressions respectively in \eqref{4.1},
\eqref{4.2} and \eqref{4.3}. We get successively
\begin{align}
<e^{i\theta }\mid z;\gamma ,m>&=\left( \pi ^{-1}\left( \gamma
-2m\right) \left( 1-\left| z\right| ^{2}\right) ^{-1-\gamma }\right)
^{-\frac{1}{2}} \nonumber \\
&\times \sum_{n=0}^{+\infty }\left( -1\right) ^{n}\left( \frac{\gamma -2m}{%
\pi }\right) ^{\frac{1}{2}}\left( \frac{n!\Gamma \left( \gamma -m+1\right) }{%
m!\Gamma \left( \gamma -2m+n+1\right) }\right) ^{\frac{1}{2}}  \label{4.6}
\\
&\times \left( 1-\left| z\right| ^{2}\right) ^{-m}\overline{z}%
^{m-n}P_{n}^{\left( m-n,\gamma -2m\right) }\left( 1-2\left| z\right|
^{2}\right)  \notag \\
&\times \frac{\sqrt{\left( \gamma ^{\prime }+1\right) _{n}}}{\sqrt{n!}}%
\; {_2F_1}\left( -n,\frac{\gamma ^{\prime }}{2}+1,\gamma ^{\prime
}+1\big | 1-e^{i\theta }\right).  \nonumber
\end{align}
Then, Eq. \eqref{4.6} reduces to
\begin{align}
<e^{i\theta }\mid z;\gamma ,m>&=\frac{\sqrt{\Gamma \left( \gamma -m+1\right) }%
}{\sqrt{m!}}\left( 1-\left| z\right| ^{2}\right) ^{\frac{\gamma +1}{2}-m}%
\overline{z}^{m} \nonumber
\\&\times \sum_{n=0}^{+\infty }\frac{\left( -1\right) ^{n}\overline{z}^{-n}}{%
\sqrt{\Gamma \left( \gamma -2m+n+1\right) }}P_{n}^{\left( m-n,\gamma
-2m\right) }\left( 1-2\left| z\right| ^{2}\right)   \label{4.7}
\\&\times \frac{\sqrt{\Gamma \left( \gamma -2m+n+1\right) }}{\sqrt{\Gamma
\left( \gamma -2m+1\right) }} \; {_2F_1}\left( -n,\frac{\gamma ^{\prime }}{2}%
+1,\gamma ^{\prime }+1 \big | 1-e^{i\theta }\right). \nonumber
\end{align}
Eq. \eqref{4.7}  can also be written as
\begin{align}
<e^{i\theta }\mid z;\gamma ,m>&=\left( \frac{\Gamma \left( \gamma
-m+1\right) }{m!\Gamma \left( \gamma -2m+1\right) }\right) ^{\frac{1}{2}%
}\left( 1-\left| z\right| ^{2}\right) ^{\frac{\gamma +1}{2}-m}\overline{z}%
^{m}  \label{4.8} \\
&\times \sum_{n=0}^{+\infty }\left( -1\right) ^{n}\overline{z}%
^{-n}P_{n}^{\left( m-n,\gamma -2m\right) }\left( 1-2\left| z\right|
^{2}\right) \; {_2F_1}\left( -n,\frac{\gamma ^{\prime }}{2}+1,\gamma
^{\prime }+1\big | 1-e^{i\theta }\right) .  \notag
\end{align}
Now, we set
\begin{equation}
<e^{i\theta }\mid z;\gamma ,m>=\left( \frac{\Gamma \left( \gamma -m+1\right)
}{m!\Gamma \left( \gamma -2m+1\right) }\right) ^{\frac{1}{2}}\left( 1-\left|
z\right| ^{2}\right) ^{\frac{\gamma +1}{2}-m}\overline{z}^{m}\frak{G}%
_{\gamma ,m}  ,\label{4.9}
\end{equation}
where
\begin{equation}
\frak{G}_{\gamma ,m}:=\sum_{n=0}^{+\infty }\left( -1\right) ^{n}\overline{z}%
^{-n}P_{n}^{\left( m-n,\gamma -2m\right) }\left( 1-2\left| z\right|
^{2}\right) \; {_2F_1}\left( -n,\frac{\gamma ^{\prime }}{2}+1,\gamma
^{\prime }+1\big | 1-e^{i\theta }\right) .  \label{4.10}
\end{equation}
Next, making use of the known fact on Jacobi polynomials :
\begin{equation}
P_{n}^{\left( m-n,\gamma -2m\right) }\left( 1-2\left| z\right| ^{2}\right)
=\left( -1\right) ^{n}P_{n}^{\left( \gamma -2m,m-n\right) }\left( 2\left|
z\right| ^{2}-1\right)   \label{4.11}
\end{equation}
and writing the polynomial in right hand side\ of \eqref{4.11} as
(\cite[p.999]{2}):
\begin{equation}
P_{n}^{\left( \gamma ^{\prime },m-n\right) }\left( u\right) =\left(
\begin{array}{c}
n+\gamma ^{\prime } \\
n
\end{array}
\right) \left( \frac{1+u}{2}\right) ^{n}\; {_2F_1}\left( -n,-m,\gamma
^{\prime }+1\bigg | \frac{u-1}{u+1}\right)   \label{4.12}
\end{equation}
for $u=2\left| z\right| ^{2}-1,$ $\gamma ^{\prime }=\gamma -2m,$ we get that
\begin{equation}
P_{n}^{\left( \gamma ^{\prime },m-n\right) }\left( 2\left| z\right|
^{2}-1\right) =\frac{\left( \gamma ^{\prime }+1\right) _{n}}{n!}\left|
z\right| ^{2n}\; {_2F_1}\left( -n,-m,\gamma ^{\prime }+1\bigg | \frac{\left|
z\right| ^{2}-1}{\left| z\right| ^{2}}\right)   \label{4.13}
\end{equation}
So that Eq. \eqref{4.10} takes the form
\begin{eqnarray}
\frak{G}_{\gamma ,m} &=&\sum_{n=0}^{+\infty }\overline{z}^{-n}P_{n}^{\left(
\gamma -2m,m-n\right) }\left( 2\left| z\right| ^{2}-1\right)
\; {_2F_1}\left( -n,\frac{\gamma ^{\prime }}{2}+1,\gamma ^{\prime
}+1;1-e^{i\theta }\right)   \label{4.14} \\
&=&\sum_{n=0}^{+\infty }\frac{\left( \gamma ^{\prime }+1\right) _{n}}{n!}%
\; {_2F_1}\left( -n,-m,\gamma ^{\prime }+1\bigg | \frac{\left| z\right| ^{2}-1}{%
\left| z\right| ^{2}}\right) \; {_2F_1}\left( -n,\frac{\gamma ^{\prime }}{2}%
+1,\gamma ^{\prime }+1\big | 1-e^{i\theta }\right) z^{n}  \notag
\end{eqnarray}
Now, we apply the generating formula (\cite[p.1037]{6}:
\begin{align}
\sum_{n=0}^{+\infty }\frac{\left( \beta \right) _{n}}{n!} &\; {_2F_1}\left(
-n,\alpha ,\beta ;x\right) \;  {_2F_1}\left( -n,\delta ,\beta |y\right) z^{n}
\label{4.15} \\
&=\frac{\left( 1-z\right) ^{\alpha +\delta -\beta }}{\left( 1+\left(
x-1\right) z\right) ^{\alpha }\left( 1+\left( y-1\right) z\right) ^{\delta }}%
\; {_2F_1}\left( \alpha ,\delta ,\beta \bigg | \frac{xyz}{\left( 1+\left(
x-1\right) z\right) (1+\left( y-1\right) z}\right)   \notag
\end{align}
for the parameters
\begin{equation}
\beta =\gamma ^{\prime }+1,\text{ \ \ }\alpha =-m,\text{ \ \ }\delta =\frac{%
\gamma ^{\prime }}{2}+1,x=\frac{\left| z\right| ^{2}-1}{\left| z\right| ^{2}}%
,\text{ \ }y=1-e^{i\theta } . \label{4.16}
\end{equation}
This gives
\begin{equation}
\frak{G}_{\gamma ,m}=\frac{\left( \overline{z}-1\right) ^{m}\left(
1-ze^{i\theta }\right) ^{-\frac{\gamma }{2}-1+m}}{\overline{z}^{m}\left(
1-z\right) ^{\frac{\gamma }{2}}}\; {_2F_1}\left(
\begin{array}{c}
-m,\frac{\gamma }{2}-m+1 \\
1+\gamma -2m
\end{array}
\bigg |  \frac{\left( 1-\left| z\right| ^{2}\right) \left( 1-e^{i\theta }\right)
}{\left( 1-\overline{z}\right) \left( 1-ze^{i\theta }\right) }\right).
\label{4.17}
\end{equation}
Return back to \eqref{4.9} and replacing $\frak{G}_{\gamma ,m}$ by
its expression in \eqref{4.17}  we arrive at the expression
\eqref{4.5}. This ends the proof. \fin \\

Now, let $\phi \in L^{2}(\mathbf{S}^{1},d\sigma _{\left( \gamma
-2m\right) }) $. Using the point (3) of \ the formalism in Section
2, we can define
\begin{align}
\mathcal{B}_{\gamma ,m}^{J}\left[ \phi \right] (z) &=\left(
K_{\gamma ,m}\left( z,z\right) \right) ^{\frac{1}{2}}\left\langle
\phi ,\mid z;\gamma ,m>\right\rangle _{\mathcal{H}}  \label{4.18}
\\
&=\left( \pi ^{-1}\left( \gamma -2m\right) \left( 1-\left| z\right|
^{2}\right) ^{-1-\gamma }\right) ^{\frac{1}{2}}\int_{0}^{2\pi
}<e^{i\theta }\mid z;\gamma ,m>\phi \left( e^{i\theta }\right)
d\sigma _{\gamma }\left( \theta \right).   \label{4.19}
\end{align}
Next, writing the closed form of the wave function in proposition
\eqref{4.1} in terms of Jacobi polynomials we can state the
following result.

\begin{theorem} \label{Thm4.1}
Let $\gamma >0$ and $m=0,1,..,\left[ \frac{%
\gamma }{2}\right] .$  Then, the coherent state transform associated
with the coherent states in \eqref{4.4} is the isometry $B_{\gamma ,m}^{J}:L^{2}(%
\mathbf{S}^{1},d\sigma _{\left( \gamma -2m\right) })\rightarrow
A_{m}^{\gamma }\left( \mathbb{D}\right) $  defined by
\begin{eqnarray}
\mathcal{B}_{\gamma ,m}^{J}\left[ \phi \right] (z)  &=&\frac{%
\left( \frac{\Gamma \left( \gamma +1-m\right) }{\pi m!\Gamma \left( \gamma
-2m\right) }\right) ^{\frac{1}{2}}}{\left( 1-z\right) ^{\frac{\gamma }{2}}}%
\int_{0}^{2\pi }\frac{1}{\left( 1-e^{i\theta }z\right)
^{\frac{\gamma }{2}+1}}\left( \frac{\left( \overline{z}-1\right)
\left( 1-ze^{i\theta }\right) }{\left( 1-\left| z\right| ^{2}\right)
}\right) ^{m}  \label{4.20}
\\
&&\times {_2F_1}\left(
\begin{array}{c}
-m,\frac{\gamma }{2}-m+1 \\
1+\gamma -2m
\end{array}
\bigg |  \frac{\left( 1-e^{i\theta }\right) \left( 1-\left| z\right| ^{2}\right)
}{\left( 1-\overline{z}\right) \left( 1-e^{i\theta }z\right) }\right)
\overline{\phi \left( e^{i\theta }\right) }d\sigma _{\left( \gamma
-2m\right) }\left( \theta \right). \nonumber
\end{eqnarray}
\end{theorem}

\begin{definition} \label{Def4.11}  The coherent state transform
 $B_{\gamma ,m}^{J}$  in \eqref{4.20} will be called the circular
Bargmann transform attached to the  $m$th hyperbolic Landau level.
\end{definition}

\begin{remark} \label{Rem4.1} Note that\ when $m=0$, the transform in
\eqref{4.20} reduces to the first one in \eqref{3.20}. i.e.,
$\mathcal{B}_{\gamma ,0}^{J}=\mathcal{B}_{\gamma }^{J}$.
\end{remark}

\begin{remark} \label{Rem4.2} In view of the projective mapping of the unit circle $%
\mathbf{S}^{1}$onto the interval $[-1,1]$, it is possible to write
the action of the circular Bargmann transform $\mathcal{B}_{\gamma
,m}^{J}$ on square integrable functions on $[-1,1]$ with respect to
an appropriate measure.
\end{remark}

\end{document}